\documentclass[12pt]{article}
 \usepackage[cp1251,koi8-r]{inputenc}
 \usepackage[english,russian]{babel}
 \usepackage{graphicx}
 \usepackage{amsfonts, amsmath,amssymb, cite}
 \inputencoding{cp1251}
 \fontencoding{T2A}\selectfont
 \numberwithin{equation}{section}
 \renewenvironment{abstract}
 {\begin{center} {\ } \end{center} \quotation}
 {\endquotation}

\newcommand\Pd{\tfrac{1}{2}}
\newcommand\e{{\mathrm{e}}}
\newcommand\n{{\mathrm{n}}}
\newcommand\ex{{\mathrm{ex}}}
\newcommand\BEC{{\mathrm{BEC}}}
\newcommand\BOX{{\mathrm{box}}}
\newcommand\Trap{{\mathrm{trap}}}

\newcommand\Erfc{{\mathrm{Erfc\,}}}
\newcommand\Ima{{\mathrm{Im}}}
\newcommand\Li{{\mathrm{Li}}}

\newcommand\be{\begin{equation}}
\newcommand\ee{\end{equation}}
\newcommand\bee{\begin{eqnarray}}
\newcommand\eee{\end{eqnarray}}

\begin{document}

 \title{On the theory of Bose-condensate fluctuations in systems of finite size}{}%
\author{A. I. Bugrij and V. M. Loktev  \\
{\small{\it N. N. Bogolyubov Institute for Theoretical Physics of the NAS of
Ukraine, }}\\
 {\small{\it  ul. Metrologicheskiaya 14-b, Kiev 03143, Ukraine }}\\
 {\small E-mail: abugrij@bitp.kiev.ua}\\
 {\small \ \hphantom{E-mail:}vloktev@bitp.kiev.ua}}
 \date{}
 \maketitle

 \begin{abstract}
 {\bf Abstract}

An asymptotic expansion for the grand partition function of an ideal Bose
gas is obtained for the canonical ensemble with arbitrary number of
particles. It is shown that the expressions found are valid at all
temperatures, including the critical region. A comparison of the
asymptotic formulas for fluctuations of the Bose condensate with exact
ones is carried out and their quantitative agreement is established.

\end{abstract}

\section*{Introduction}
The statistical mechanics of an ideal Bose gas \cite{KH, LL} is one of the
simplest while at the same time fundamental subject areas of theoretical
physics. The most impressive result of the theory is the remarkable
phenomenon of Bose-Einstein condensation (BEC), which was in fact
predicted by Einstein \cite{1} the accumulation of an unlimited number of
noninteracting (and this is also nontrivial) particles in their quantum
ground state. We note that BEC turned out to be the first exactly solvable
model of a phase transition and is a demonstration of the possibility of
quantum behavior on macroscopic scales. Unfortunately, there are no
substances in nature that satisfy the conditions of BEC in thermodynamic
equilibrium. Nevertheless, the idea of the the Bose condensate has turned
out to be very useful for understanding the essence of such remarkable
physical phenomena as superfluidity and superconductivity and has been
fruitful in various conceptual specu\-la\-ti\-ons in condensed matter
physics and quantum field theory.

By the end of the last century experimental technique had been perfected so much that
it became possible to restrain a cloud of polarized atoms of hydrogen or alkali metals in a
magnetic trap long enough to cool it to very low temperatures ($\sim20~n{\rm K}\ldots \sim1~\mu{\rm K}$)
\cite{2, 3, 4, 5} as a
result, a large number of atoms ($\sim10^{3}$ in the case of rubidium and $\sim10^{5}$  for sodium) have
been collected in the lowest-energy quantum state. Although it is understood that it is not
in a state of thermodynamic equilibrium, this is commonly regarded as the experimental
preparation of a Bose condensate.

If one has in mind not a strict thermal equilibrium but a quasi-equilibrium, then it is
entirely acceptable to consider a collective of some quasiparticles whose mutual interaction
is also weak, as a rule, to be a suitable object for experimental investigation of BEC. As
we know, the condition for thermal equilibrium here requires that the chemical potential
vanish ($\mu=0$), while the BEC regime corresponds to an increase of the chemical potential
to the lowest energy level ($\mu\to\varepsilon_{0}$). In the case when $\varepsilon_{0}$ is small, an equilibrium gas of
quasiparticles is found at the threshold of Bose condensation. However, in quasiequilibrium
(e.g., under the influence of the external pump creating the quasiparticles) it is possible in
principle to control the chemical potential, increasing it to values close to $\varepsilon_{0}$ and thereby
satisfying the necessary conditions for BEC.

One can give examples of phenomena from everyday life in which the BEC of
quasipar\-tic\-les in essence takes place. For example, from the
standpoint of the quantum theory of solids, such processes as the ringing
of various vessels, the sounding of tuning forks, etc., that are just the
BEC of phonons ``pumped'' by the striking impact or, in other words, the
preparation of a phonon Bose condensate, which, in turn, generates
coherent sound emission. Remarkably, all this occurs at normal
temperatures and so does not require any special experimental
contrivances. At the same time, serious efforts continue to be made in the
investigation of the BEC of such quasiparticles as excitons and biexcitons
in semiconductors \cite{6} and magnons in certain classes of quantum
magnets \cite{7}. Perhaps the most impressive are the recent results of
Demokritov and co-authors \cite{8} with microfilms of yttrium iron garnet.
Their Mandelstam-Brillouin scattering experiments cleared revealed a
resonance peak in the spectral density of the distribution of magnons over
states as a function of frequency in the vicinity of the minimum energy of
the corresponding energy spectrum upon an increase of pumping below a
certain threshold value. These experiments might be described as having
created an analog of the ``magnon tuning fork'', and at high (room)
temperatures.

Although it can be said that our present understanding of the various properties of Bose
condensates of particles and quasiparticles is adequate, a number of unanswered questions
remain, in particular, in regard to fluctuations of some observable quantities. This pertains
primarily to systems of finite size or with a finite number of condensing particles, when the
very concept of thermodynamic limit becomes problematical. Here it should be mentioned
that the samples used in experiments on BEC are not only finite but usually small in size.
If the spatial dimensions of the system are finite, then the statistical mechanics of the
Bose gas is substantially complicated. In particular, the concept of equivalence of canonical
ensembles loses meaning. We note here that in the BEC regime, inequivalence of ensembles
is manifested even in the thermodynamic limit. Some finite-size effects for a Bose gas in a
box were discussed in \cite{9}. The ``poor'' behavior of fluctuations of the condensed particles
described in the framework of the grand canonical ensemble (GCE) prompted the authors
of \cite{9} to draw the radical conclusion that this ensemble is unsuitable for describing
any real physical system undergoing BEC. We add that the entropy also behaves ``poorly''
in the GCE - as the temperature goes to zero it does not go to zero in accordance with
the Nernst theorem but, on the contrary, diverges (logarithmically) when the number of
particles $N\to\infty$. However, the GCE, because of its simplicity, is extremely convenient
for concrete calculations; moreover, the behavior of such quantities as, e.g., the number of
particles in the ground state, $N_{0}$, or the specific heat does not differ in calculations using
different ensembles.

    The experimental study of BEC in traps has revived the theoretical research on different
aspects of the statistical mechanics of Bose systems \cite{10}--\cite{Rez}. This is primarily because of the
fact that until then attention had mainly been devoted to a spatially homogeneous gas
found in a certain volume. In the traps that were actually used the gas is inhomogeneous,
and therefore the results obtained previously for a homogeneous Bose gas in a box can
be reproduced for a trap by considering it to be a potential well with a harmonic law of
spatial confinement, i.e., basically reducing the problem to a system of $N$ oscillators.

Their
partition function in the GCE (the grand partition function) is trivial to calculate. In the
canonical (CE) and microcanonical (MCE) ensembles the corresponding partition functions
are expressed in terms of contour integrals of the grand partition function. Therefore for
not too large a number of particles and not very high energy, if one is talking about the
MCE, the partition function of the Bose system can be calculated to any desired accuracy
by the residue theorem or numerical integration. As to the analytical calculations of the
partition function in the CE and MCE, one usually uses the saddle-point (steepest descent)
method. However, as was shown in \cite{9}, for example, that method is inapplicable in the
most interesting region - the neighborhood of the BEC, where the corrections to the main
contribution do not fall off with increasing number of particles in the system.

A number of calculations have been done to investigate the thermodynamic properties of
finite Bose systems in different statistical ensembles. In particular, in \cite{10} the difference
between the behavior of the number of condensate particles $N_{0}$ and also their fluctuations,
calculated for a harmonic trap with the use of the GCE or CE when the total number of
particles varied in the range $10^{2}\leq N\leq10^{6}$. In \cite{11}, which was devoted to a comparison
of the results of an exact calculation in the MCE write approximate results obtained by
the saddle-point method, it was noted that these results differ substantially precisely in the
neighborhood of the BEC point.

Nevertheless, to this day there is no convenient analytical representation
for the partition function in the CE and MCE which would correspond well
enough to the BEC regime. The goal of the present study is to remedy this.
In Sec. I we introduce the necessary notation and definitions and also
obtain an analytical expression for the average number of particles
$N_{0}$ on the ground level under the condition $N_{0}\gg1$. In the Sec.
II we calculate the partition function in the CE by the saddle-point
method with the first correction taken into account. In Sec. III we
analyze why the domain of applicability of the saddle-point method is
limited to temperatures $T>T_{BEC}$; here by isolating the singularity
corresponding to the ground level, we derive an expression valid for
$T<T_{BEC}$ as well. In Sec. IV we propose a method of asymptotic
expansion of the partition function in the CE in inverse powers of the
number of particles, which works both above and below the BEC temperature.
On the basis of the representation obtained, we calculate the fluctuations
of the Bose condensate and demonstrate the quantitative agreement with the
exact result down to very small values of the number of particles in the
system. In the Conclusion we discuss the BEC temperature for a harmonic
trap and a box and also the difference of the mathematical mechanism of
formation of the critical point in the GCE and CE.

 \section{Grand Canonical Ensemble}
 A stationary quantum system consisting of $N$ noninteracting particles is known to be
completely characterized by the configuration
$[\n]=\{\n_{0},\n_{1},\n_{2},\ldots\}$, where $\n_{k}=0,1,2,\ldots$ is the
number of particles in the $k$th quantum states ($(k=0,1,2,\ldots)$
According to the precepts of statistical mechanics, the (time) average of
an observable quantity $A$ in a nonstationary system coincides with to the
average over an ensemble of stationary systems. An ensemble is determined
by the distribution function $\rho[\n]$. Then
 \be\label{1}
\overline{A}=Z^{-1}\sum_{[\n]}\rho[\n]A[\n]\,,\ee
where the normalizing coefficient $Z$ (the partition function) has the form
 \be\label{2}
Z=\sum_{[\n]}\rho[\n]\,.\ee
In the description of an ideal Bose gas one generally uses the GCE, with a distribution
function
 \be\label{3}
\rho[\n]=\e^{\sum\limits_{k}\n_{k}(\mu-\varepsilon_{k})/T}\,,\ee where $T$
is the temperature, $\mu$ is the chemical potential, and $\varepsilon_{k}$
is the single-particle energy of the $k$th state. Here and below we have
set Boltzmann's constant $k_{B}=1$. Since $\rho[\n]$ (\ref{3}) is
factorized with respect to the dependence on the occupation numbers
$\n_{k}$, the summation over configurations $[\n]$ is trivial to do. In
particular, the average number of particles in the $k$th state
 \be\label{4}
\overline{\n}_{k}=Z^{-1}\sum_{[\n]}\rho[\n]\n_{k}=
\frac{\sum\limits_{\n=0}^{\infty}\n\,\e^{\n(\mu-\varepsilon_{k})/T}}
{\sum\limits_{\n=0}^{\infty}\e^{\n(\mu-\varepsilon_{k})/T}}=
\frac{1}{\e^{(\varepsilon_{k}-\mu)/T}-1}\,.\ee
The function on the right-hand side of (\ref{4}) specifies the average occupation number and,
hence, is a constituent element of the expressions for the majority of thermodynamic quantities
(and not only in the GCE), and it is deviate to use a special notation for it:
 \be\label{5} n_{k}=\frac{1}{\e^{(\varepsilon_{k}-\mu)/T}-1}=
\frac{1}{\e^{(\varepsilon_{k}-\varepsilon_{0})/T}(1+n_{0}^{-1})-1}\,.\ee
With the use of (5) the partition function (\ref{2}) is expressed as the product
(\ref{5})
\be\label{5f}Z=\prod_{k=0}^{\infty}(n_{k}+1)\,.\ee
The independent variables in the GCE are assumed to be $T$ and $\mu$. However, as follows from
Eq. (\ref{5}), they could be considered to be the temperature and the number of particles $n_{0}$  in
the (ground) state with the lowest energy $\varepsilon_{0}$, which facilitates the analysis of the different
regimes of the GCE.

We write the average value of the total number of particles in the form
\be\label{6}
\overline{N}=\sum_{k=0}^{\infty}\overline{\n}_{k}=n_{0}+N_{\ex}(n_{0,}T)\,,\ee
where   \be\label{7} N_{\ex}(n_{0},T)\equiv
N_{\ex}=\sum_{k=1}^{\infty}n_{k}\ee
is the average number of particles in the excited states. If the value of $\overline{N}$
is fixed, then all
the numbers $n_{k}$ except $n_{0}$ fall with decreasing temperature ($n_{k\neq0}=0$ at $T=0$), and $n_{0}$
grows to $n_{0}=\overline{N}$
at $T=0$. We denote by $\widetilde{n}_{k}$ the maximum possible value of the average
occupation number at a given temperature, i.e., the value of $n_{k}$ of Eq. (\ref{5}) for $n_{0}\to\infty$ (or,
equivalently, $\mu=\varepsilon_{0}$):
\be\label{8}
\widetilde{n}_{k}=\frac{1}{\e^{(\varepsilon_{k}-\varepsilon_{0})/T}-1}\,.\ee
Then for $n_{0}\gg1$ for the number $n_{k}$ specified by Eq. (\ref{5}) one can limit consideration to the
expansion
 \be\label{9}
n_{k}\simeq\widetilde{n}_{k}-\frac{\widetilde{n}_{k}(\widetilde{n}_{k}+1)}{n_{0}}\,,\ee
which, in turn, reduces Eq. (\ref{6}) for $n_{0}$ to a simple quadratic equation:
  \be\label{10}
\overline{N}=n_{0}+\widetilde{N}_{\ex}-\frac{\delta\widetilde{N}_{\ex}^2}{n_{0}}\,,\ee
in which \be\label{11}
\widetilde{N}_{\ex}=\sum_{k=1}^{\infty}\widetilde{n}_{k}\,,\qquad
\delta\widetilde{N}_{\ex}^2=\sum_{k=1}^{\infty}\widetilde{n}_{k}(\widetilde{n}_{k}+1)\ee
is the maximum possible number of particles in excited states and its
mean-square fluc\-tu\-a\-ti\-on. We note that the quantities marked with a
tilde are functions of temperature only. The solution of equations
(\ref{10}) with respect to $n_{0}$ is denoted as
  \be\label{12}
N_{0}(T)=\frac{1}{2}\biggl(\overline{N}-\widetilde{N}_{\ex}+
\sqrt{(\overline{N}-\widetilde{N}_{\ex})^{2}+4\delta\widetilde{N}_{\ex}^2}\,\biggr)\,,\ee
which we shall call the Bose condensate. This terminology is conditional in the sense that
one is considering a problem outside the thermodynamic limit, with a finite total number of
particles $\overline{N}$. It follows from the definitions (\ref{8}) and (\ref{11}) that $\widetilde{N}_{\ex}(T)$ and $\delta\widetilde{N}_{\ex}^2(T)$ are
monotonically increasing functions of temperature. We denote by $T_{c}$ the temperature at
which $\widetilde{N}_{\ex}$ is equal to $\overline{N}$, which corresponds to
 \be\label{13}
\widetilde{N}_{\ex}(T_{c})=\overline{N}\,.\ee
If $\delta\widetilde{N}_{\ex}^2\ll(\overline{N}-\widetilde{N}_{\ex})^{2}$, and this holds for $T\neq T_{c}$ and $\overline{N}\gg1$, then the behavior of solution
(\ref{12}) in the limit $N\to\infty$ acquires a stepped character. For different temperature regions,
both below $T_{c}$, where $\overline{N}>\widetilde{N}_{\ex}$, and above $T_{c}$, where $\overline{N}<\widetilde{N}_{\ex}$, the asymptotic behavior of
$N_{0}(T)$ at large but finite $\overline{N}$ has the simple form
\be\label{14}
N_{0}(T)=\left\{\begin{array}{ll}\overline{N}-\widetilde{N}_{\ex},&T<T_{c},\\
\delta\widetilde{N}_{\ex},&T=T_{c},\\
\delta\widetilde{N}_{\ex}^2/(\widetilde{N}_{\ex}-\overline{N}),&T>T_{c}\end{array}\right..\ee
The value of $T_{c}$ at which the change of regime (the crossover) in the
behavior of $N_{0}(T)$ occurs can be regarded as a generalization of the
temperature $T_{\BEC}$ to the case of a finite number of particles in the
system. We recall that Eq. (\ref{10}) is approximate, in accordance with
the condition $n_{0}\gg1$. In the opposite case, when $n_{0}\ll1$ (the
Boltzmann limit), Eq. (\ref{6}) gives the simple depends \be\label{15}
N_{0}(T)\simeq\frac{\overline{N}}{1+Q(T)}\,,\qquad
Q(T)=\sum_{k=1}^{\infty}\e^{-(\varepsilon_{k}-\varepsilon_{0})/T}\,,\ee
which attest to the classical behavior of the Bose systems under
consideration.

Independently of the number of particles $\n_{0}$ in the condensate the factorized character of
the distribution function (\ref{3}) in the GCE is conditional upon the absence of any correlations
between particles of the Bose gas in different quantum states. This has the consequence
\be\label{16}
\overline{\n_{k}\n_{l}}=\overline{\n}_{k}\cdot\overline{\n}_{l}\,.\ee
The average of the square (and higher powers) of the number of particles in the $k$th state is
calculated in analogy with Eq. (\ref{4}):
\be\label{17}
\overline{\n_{k}^{2}}=2n_{k}^{2}+n_{k}\,,\ee
from which the mean-square deviation (or, in other words, the mean-square fluctuation) is
easily calculated and has the form
\be\label{19}
\delta\n^2_{k}=\overline{\n_{k}^{2}}-\overline{\n}_{k}\,\!\!^{2}=n_{k}(n_{k}+1).\ee
Taking Eq. (\ref{16}) into account, we write the square of the fluctuation of the total number of
particles as
\be\label{20}
\delta N^{2}=\delta\n^2_{0}+\delta N_{\ex}^2,\quad \delta
\n_{0}^{2}=n_{0}(n_{0}+1),\quad \delta
N_{\ex}^2=\sum_{k=1}^{\infty}n_{k}(n_{k}+1).\ee
For $T<T_{c}$ the value $\n_{0}\sim \overline{N}$, and the square of the fluctuation of the number of condensed
particles is
\be\label{21} \delta\n^2_{0}=n_{0}(n_{0}+1)\sim\overline{N}\,^{2}.\ee
Thus the relative fluctuation $(\delta\n^2_{0}/\overline{N})^{1/2}$ grows with increasing number of particles in the
system, and this is the basis for the widespread assertion that the fluctuations diverge below
the BEC point (see, e.g., \cite{LL}).

We note in this regard that the description of the BEC in the framework of the GCE
cannot be considered quite correct, if for no other reason that it explicitly violates the Nernst
theorem. Indeed, the entropy of the GCE is expressed in terms of the average occupation
number as
\be\label{22}
S=\sum_{k}[(n_{k}+1)\ln(n_{k}+1)-n_{k}\ln n_{k}].\ee
In the region $T\ll T_{c}$, where $n_{0}\gg1$ and $n_{k\neq0}\ll1$, it becomes equal to the entropy of the
Bose condensate:
$$S\simeq(n_{0}+1)\ln(n_{0}+1)-n_{0}\ln n_{0}\simeq\ln(n_{0}+1)+1.$$
When $T\to0$, the entropy $S\simeq\ln \overline{N}$, i.e., not only does it not go to zero but it diverges with
increasing number $\overline{N}$. As will be seen below, in the canonical ensemble there is no problem
with a divergence of the fluctuations nor with the entropy.

\section{Canonical Ensemble}
The main difference between the CE and GCE is that the total number of particles in
the CE is rigidly fixed: $N=\sum\limits_{k=0}^{\infty}\n_{k}=\n_{0}+N_{\ex}$. From this it follows directly that
\bee\label{1a}
&&\overline{\n}_{0}=N-\overline{N}_{\ex},\\\label{2a}&&(\n_{0}-\overline{\n}_{0})^{2}
=\overline{(N_{\ex}-\overline{N}_{\ex})^{2}},\eee
i.e., the fluctuation of the Bose condensate does not differ from the fluctuation of the total
number of particles in excited states, or
\be\label{3a} \delta \n_{0}=\delta
N_{\ex}.\ee

It is essential here that in the CE the average number of particles in the $k$th state is not equal
to the average occupation number, determined in Eq. (\ref{5}), and the noninteracting particles
in different quantum states (in contrast to the GCE) are correlated with each other.

The fluctuation of the Bose condensate in the CE can be estimated starting
from the following qualitative arguments. For $T>T_{c}$ the number
$\overline{\n}_{0}\ll N$. Therefore, considering the Bose condensate as a
small subsystem, one can suppose that a description of it in the framework
of the GCE is valid. Consequently, for $T>T_{c}$ the following relation
also holds in the CE [see Eq. (\ref{19})]: \be\label{4a}
\delta\n_{0}^{2}\simeq n_{0}(n_{0}+1).\ee When $T<T_{c}$, however, the
small subsystem becomes the particles above the condensate, and now the
mean-square fluctuation of their number in the CE can be described by the
expression \be\label{5a} \delta
N_{\ex}^2\simeq\sum_{k=1}^{\infty}n_{k}(n_{k}+1).\ee Doing a simple
interpolation of expressions (\ref{4a}) and (\ref{5a}) with (\ref{3a})
taken into account, we find that \be\label{6a} \delta\n^2_{0}=\delta
N_{\ex}^2=\frac{n_{0}(n_{0}+1)\sum\limits_{k=1}^{\infty}n_{k}(n_{k}+1)}
{\sum\limits_{k=0}^{\infty}n_{k}(n_{k}+1)}\,.\ee Below we compare this
phenomenological expression with the expression calculated directly in the
CE.

For this we define the distribution function in the CE [cf. Eq. (\ref{3})]
\be\label{7a}
\rho[\n]=\e^{-\sum\limits_{k}\n_{k}\varepsilon_{k}/T}\delta(N-\sum_{k}\n_{k}).\ee
The property of factorization of $\rho[\n]$ is lost because of the
presence of the $\delta$-function on the right-hand side of Eq.
(\ref{7a}), so that in comparison with the GCE the calculation of the
partition function and the averages of the observables is complicated.
Factorizability can be easily restored, however, and the summation over
configurations $[\n]$ can be reduced to a summation over independent
$\n_{k}$ if one uses the integral representation of the Kronecker
 $\delta$-function: \be\label{8a}
\delta(m)=\frac{1}{2\pi}\int\limits_{-\pi}^{\pi}dx\, \e^{ixm}.\ee Then,
substituting (\ref{8a}) into (\ref{7a}) and changing the sequence of
summation and integration, we arrive at the following representation for
the partition function: \be\label{9a}
Z=\frac{1}{2\pi}\int\limits_{-\pi}^{\pi}dx\,
\e^{ixN}\sum_{[\n]}\e^{-\sum\limits_{k}\n_{k}(\varepsilon_{k}/T+ix)}=
\frac{1}{2\pi}\int\limits_{-\pi}^{\pi}dx\,\e^{ixN+W(-ix)}\,,\ee where
$$ W(-ix)=-\sum_{k}\ln(1-\e^{-\varepsilon_{k}/T-ix}),$$
or, denoting $x$ as $i\nu$, \be\label{10a}
W(\nu)=\sum_{k=0}^{\infty}\ln(n_{k}+1), \qquad
n_{k}=\frac{1}{\e^{\varepsilon_{k}/T-\nu}-1}\,.\ee A comparison of Eqs.
(\ref{10a}) and (\ref{5f}) shows that the function $W(\nu)$ is the
logarithm of the partition function in the GCE if one sets $\nu=\mu/T$.
Thus in the CE the averages of physical quantities are just the ratios of
the corresponding integrals of the same averages found in the GCE. This,
on the one hand, establishes a definite relation between the two
statistical ensembles and, on the other, allows one to use the initial
[see Eq. (\ref{5})] definition for the number of Bose particles in each of
the states. However, in them the average in the CE can also be expressed
in terms of derivatives of the partition function. For example, in the
simplest case of a nondegenerate spectrum $\varepsilon_{k}$ we have
 \bee\label{11a}
&&\overline{\n}_{k}=Z^{-1}\sum_{[\n]}\rho[\n]\n_{k}=
-Z^{-1}T\frac{\partial}{\partial\varepsilon_{k}}\sum_{[\n]}\rho[\n]=-T\frac{\partial\ln
Z}{\partial\varepsilon_{k}}\,,\\\label{12a}
&&\delta\n_{k}^{2}=-T\frac{\partial\overline{\n}_{k}}{\partial\varepsilon_{k}}\,.\eee

At large $N$ the integral on the right-hand side of Eq. (\ref{9a}) can be evaluated by the
saddle-point method, which leads to the following asymptotic expansion:
\be\label{13a} Z=\frac{\e^{W(\nu)-\nu N}}{\sqrt{2\pi
W''(\nu)}}(1+z_{1}+z_{2}+\cdots).\ee
Here $\nu$ denotes the saddle point nearest to the origin of coordinates in the complex $x$-plane
$(x_{s}=i\nu)$, the equation for which has the form
 \be\label{14a}
W'(\nu)=\sum_{k=0}^{\infty}n_{k}=N.\ee
In the leading asymptotic approximation the logarithm of the partition function in the CE
has the following simple form:
\be\label{15a} \ln Z=W(\nu)-\nu N-\Pd\ln[2\pi W''(\nu)].\ee
The contributions $z_{j}$ in Eq. (\ref{13a}) are expressed in terms of ratios of the derivatives of the
function $W(\nu)$ of the type
\be\label{16a}
\frac{[W^{(l)}(\nu)]^{m}[W^{(k)}(\nu)]^{n}}{[W''(\nu)]^{m+n+j}}\,.\ee
If the function $W(\nu)$ and its derivatives are large, $W^{(l)}(\nu)\sim N$, and this is the case at least
in the region $T>T_{c}$ and $N\gg1$, then the ratios (\ref{16a}) and the contributions $z_{j}$ have order
of smallness $O(N^{-j})$. Thus for the first correction we find:
\be\label{17a}
z_{1}=\frac{W^{(4)}(\nu)}{8[W''(\nu)]^{2}}-
\frac{5[W^{'''}(\nu)]^{2}}{24[W''(\nu)]^{3}}=O(N^{-1}).\ee

We take the derivative of the occupation number (\ref{10a}) with respect to $\varepsilon_{l}$:
\be\label{18a} T\frac{\partial n_{k}}{\partial\varepsilon_{l}}=n_{k}(n_{k}
+1)\biggl(T\frac{\partial\nu}{\partial\varepsilon_{l}}-\delta_{kl}\biggr)\,.\ee
The derivative of the saddle point $\nu$ with respect to $\varepsilon_{l}$ is evaluated by differentiating Eq.
(\ref{14a}):
\be\label{19a}
T\frac{\partial\nu}{\partial\varepsilon_{l}}=\frac{n_{l}(n_{l}+1)}{W''(\nu)}\,.\ee
Now with the aid of Eqs. (\ref{11a}), (\ref{15a}), (\ref{18a}), and (\ref{19a}) we find the average value of the number
of particles in the $k$th state,
 \be\label{20a}
\overline{\n}_{k}=n_{k}-\frac{n_{k}(n_{k}+1)}{2W''(\nu)}\left(2n_{k}+1-
\frac{W^{'''}(\nu)}{W''(\nu)}\right)\,.\ee
In particular, for the average number of particles in the ground state (\ref{11a}) we obtain
 \be\label{21a}
\overline{\n}_{0}=n_{0}-\frac{n_{0}(n_{0}+1)}{2W''(\nu)^{2}}
\bigl[(2n_{0}+1)V''(\nu)-V^{'''}(\nu)\bigr],\ee
where $V(\nu)$ denotes the sum over only the excited states,
\be\label{22a}
V(\nu)=W(\nu)-\ln(n_{0}+1)=\sum_{k=1}^{\infty}\ln(n_{k}+1).\ee
finally, differentiating $\overline{\n}_{0}$ with respect to $\varepsilon_{0}$ [see Eq.~(\ref{12a})], we arrive at an expression for the mean-square fluctuation of the Bose condensate:
 \be\label{23a}
\delta\n_{0}^{2}=\delta_{1}+\delta_{2}+\delta_{3}\,,\ee
with the leading contributions
\bee \label{24a}\delta_{1}&=&\frac{n_{0}(n_{0}+1)V''(\nu)}
{W''(\nu)}\,,\\\label{25a}\delta_{2}&=
&\frac{\delta_{1}}{2W''(\nu)^{2}}\biggl\{(2n_{0}+1)V^{'''}(\nu)
-(6n_{0}^{2}+6n_{0}+1)V''(\nu)+\\\nonumber
&+&\frac{2n_{0}(n_{0}+1)(2n_{0}+1)}{W''(\nu)}\bigl[(2n_{0}+1)
V''(\nu)-V''(\nu)\bigr]\biggr\}\,,\\
\label{26a}\delta_{3}&=&\frac{n_{0}^{2}(n_{0}+1)^{2}}{2W''(\nu)^{3}}
\biggl\{(2n_{0}+1)V^{'''}(\nu)-V^{(4)}(\nu)-\\\nonumber&-&\frac{2V^{'''}(\nu)}{W''(\nu)}
\bigl[(2n_{0}+1)V''(\nu)-V^{'''}(\nu)\bigr]\biggr\}\,.\eee
In the temperature region $T>T_{c}$ the condensate is dilute, $n_{0}\ll N$, and, accordingly,
 \be\label{27a} \delta_{1}\simeq n_{0}(n_{0}+1),\quad
\delta_{2}\sim\frac{\delta_{1}}{N}\,,\quad
\delta_{3}\sim\frac{\delta_{1}^{2}}{N^{2}}\,.\ee
We note that the phenomenological formula (\ref{6a}) for the fluctuation of the Bose condensate
coincides with the leading asymptotic contribution $\delta_{1}$ (\ref{24a}).

An exact expression for the partition function is given by the single integral (\ref{9a}), which
for not too large $N$ is easily found numerically. It is interesting here to compare the exact
expression for the fluctuations with its asymptotic behavior given by formulas (\ref{23a})--(\ref{26a}).
Such a comparison, however, is impossible to do in general form, since the quantitative calculations
require specifying the explicit form of the function $W(\nu)$, which, in turn, depends
on the concrete form of the energy spectrum $\varepsilon_{k}$. Let us find it for the case of alkali metal
atoms in magnetic traps.

Experiments on cooling of a large number of alkali metal atoms $(N\simeq10^{3}\ldots10^{4})$ are
interpreted as the experimental realization of BEC. The particles are confined in the traps
by a potential $v(\mathbf{r})$, the exact dependence of which on the distance $\mathbf{r}$ is, gives speaking,
unknown, but for theoretical analysis usually a quadratic (harmonic) approximation is used.
As a result, the problem of BEC reduces, as we have said, to a calculation of the partition
function of a system of linear oscillators. The spectrum $\varepsilon_{l}$ and the spectral density $g_{l}$
(coefficient of degeneracy) of the three-dimensional isotropic oscillator has the simple form
\bee\label{28a}
&&\varepsilon_{l}=\hbar\omega\bigl(l+\tfrac{3}{2}\bigr)\,,\quad
g_{l}=\Pd(l+1)(l+2),\quad l=0,1,2,\ldots\ ,\\\nonumber&&g(\varepsilon)=
\frac{1}{2}\biggl(\frac{\varepsilon^{2}}{\hbar^{2}\omega^{2}}-\frac{1}{4}\biggr)\,,\quad
\varepsilon_{0}=\frac{3}{2}\hbar\omega.\eee
We point out that in Eq. (\ref{28a}) the index $l$ enumerates the energy levels and not quantum
states, which are enumerated by the index k introduced previously. At high temperatures
$(T\gg \varepsilon_{0})$ the series expressions for the function $W(\nu)$ and its derivatives converge slowly.
It is shown in the Appendix how to improve their convergence and to obtain expressions
convenient for numerical calculations.

The equation for the crossover point in the CE differs from the equation
(\ref{13}) in the GCE by only the replacement of $\overline{N}$ by $N$,
i.e., \be\label{28f}\widetilde{N}_{\ex}(T_{c})=N.\ee It follows from
definitions (\ref{11}) and (\ref{14a}) that
$\widetilde{N}_{\ex}=W'(\frac{\varepsilon_{0}}{T})-\widetilde{n_{0}}$.
Then, using for $W'(\nu)$ the asymptotic expansion [Eq. (\ref{21p}) in the
Appendix] for high temperatures $\tau=T/\hbar\omega\gg1$, we write the
following expansion for $\widetilde{N}_{\ex}$:
 \be\label{29a}
\widetilde{N}_{\ex}=\tau^{3}\zeta(3)+\frac{3}{2}\tau^{2}\zeta(2)+\tau\ln
\tau+O(\tau),\ee
where $\zeta(j)$ is the Riemann $\zeta$-function. From it we can find the solution of equation (\ref{28f})
that determines the crossover point $\tau_{c}$ in the form of an expansion in inverse powers of $N$.
In the leading approximation we denote this solution as
  \be\label{30a}
\tau_{\BEC}\equiv\frac{T_{\BEC}}{\hbar\omega}=\biggl[\frac{N}{\zeta(3)}\biggr]^{1/3}\,.\ee
Now, knowing equation (\ref{30a}), we find for $\tau_{c}$ from (\ref{28f}) and (\ref{29a})
 \be\label{31a}
\frac{\tau_{c}}{\tau_{\BEC}}=\frac{T_{c}}{T_{\BEC}}=
1-\frac{1}{\zeta(3)\tau_{\BEC}}\biggl[\frac{\zeta(2)}{2}+\frac{\ln\tau_{\BEC}}
{3\tau_{\BEC}}\biggr]+O(\tau_{\BEC}^{-2}).\ee

It is seen from expressions (\ref{30a}) and (\ref{31a}) that the crossover temperature $T_{c}$ is below the
condensation temperature $T_{\BEC}$ for a Bose gas in a trap. We note that for a Bose gas in a
box the situation is the opposite, $T_{c}>T_{\BEC}$.

Expressions (\ref{30a}) and (\ref{31a}) with the known numerical values of the $\zeta$-function in them
easily convince one that even for a number of particles of the order of $10^{3}$ the condensation
temperature $T_{c}$ is only 6 times greater than the ground-state energy $\varepsilon_{0}$. This is a direct
indication that the condensation phenomenon observed in the experiments mentioned is
of a microscopic (or, in any case, mesoscopic) rather than macroscopic character. This
casts doubt on whether the condensation of several thousand particles can be regarded
unambiguously as BEC, the main feature of which, strictly speaking, is the appearance and
manifestation of quantum properties in macroscopic phenomena or objects.

Without denying, of course, the presence of the phenomenon of BEC itself in magnetic
traps, we would nevertheless like to say that, in our view, the results set forth in this Section
are evidence that the condensation of alkali metal atoms observed in the experiments is more
of a nanophysical character.

In Fig.~1 we show graphs of the fluctuations of the Bose condensate in the case of their
exact calculation, $$\delta \n_{0}^{2}=T^{2}\,\frac{\partial^{2}\ln Z}{\partial
\varepsilon_{0}^{2}}$$
and their approximate calculation (\ref{23a})--(\ref{26a}). It is seen that the asymptotic expressions
represented by Eq. (\ref{23a}) and the corrections to it do not adequately reproduce the curve of
the numerical calculation.

\begin{figure}
[htb]
\begin{center}
\includegraphics[height=130mm,keepaspectratio=true]
{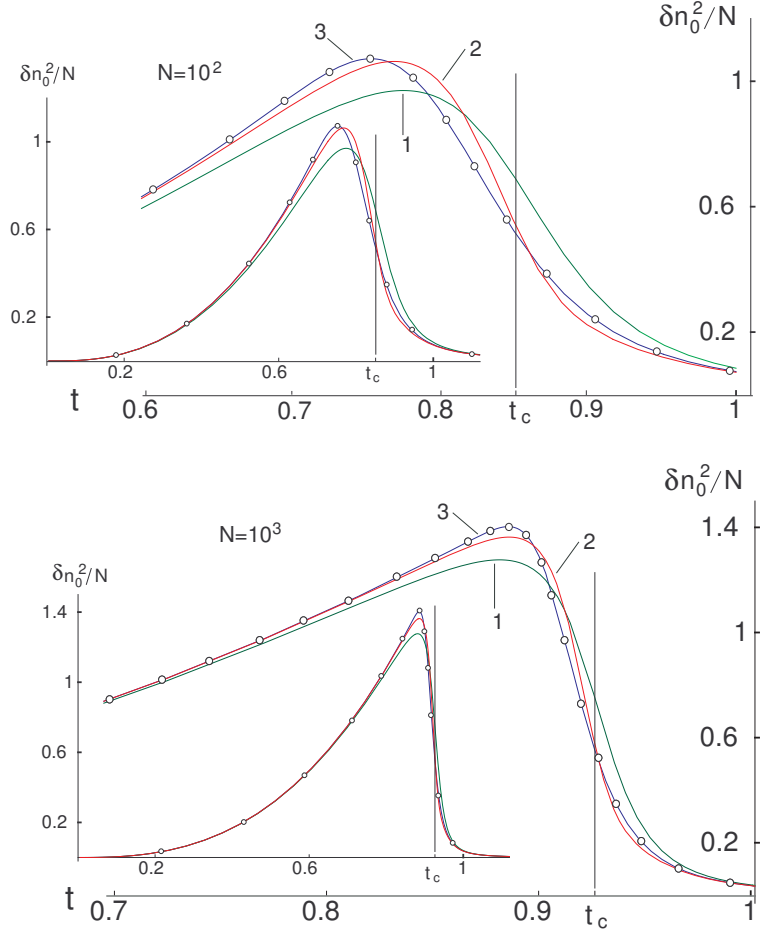} \caption{
\small
The dependence on temperature $t=T/T_{BEC}$ of
the relative fluctuation of the Bose condensate $\delta \n_{0}^{2}/N$. The
circlets denote the results of a numerical calculation of the integral in
Eq. (\ref{9a}). Curve 1 is the leading contribution to the asymptotic
expansion (\ref{23a}), curve 2 is with the next correction to Eq.
(\ref{23a}) taken into account, and curve 3 is the asymptotic expansion
(\ref{45a}), (\ref{48a}).}
\end{center} \end{figure}

\section{Bose-Einstein Condensation Region  $T<T_{c}$}

Let us consider in more detail the low-temperature region, where one can
more or less definitely talk about the presence of a Bose condensate. As
we have said (see Fig.~1), here the discrepancy between the exact and
asymptotic values of the fluctuations are significant. The reason is not
hard to understand. The fact is that in the region $T<T_{c}$ the
evaluation of integral (\ref{9a}) by the straightforward saddle-point
method does not actually give the asymptotic expansion in inverse powers
of $N$\!: the terms $z_{1}$, $z_{2},\ldots$ in Eq. (\ref{13a}) do not fall
with increasing $N$. The contributions of the ground term $w_{0}$ to the
sum $W(\nu)=\sum\limits_{k=0}^{\infty}w_{k}$ (\ref{10a}) and its
derivatives with respect to $\nu$ are: $$w_{0}=\ln(n_{0}+1),\quad
w'_{0}=n_{0},\quad w_{0}''=n_{0}(n_{0}+1), \quad
w_{0}^{'''}=n_{0}(n_{0}+1)(2n_{0}+1).$$

Since for $T<T_{c}$ the $p$th order derivative $w_{0}^{(p)}\sim n_{0}^{p}\sim N^{p}$, for the first correction, e.g., we
have
$$z_{1}\simeq\frac{1}{8}\frac{n_{0}(n_{0}+1)(6n_{0}^{2}+
6n_{0}+1)}{n_{0}^{2}(n_{0}+1)^{2}}-
\frac{5}{24}\frac{n_{0}^{2}(n_{0}+1)^{2}(2n_{0}+1)^{2}}{n_{0}^{3}(n_{0}+1)^{3}}\simeq
-\frac{1}{12}\,.$$
It is not hard to show that the other contributions $z_{j}$ in (\ref{13a}) not only do not fall  with
increasing $N$ but even grow with increasing $j$. Nevertheless, the problem of singular behavior
of the contributions corresponding to the ground state can be solved as follows.

\begin{figure}[htb]
\begin{center}
\includegraphics[height=90mm,keepaspectratio=true]
{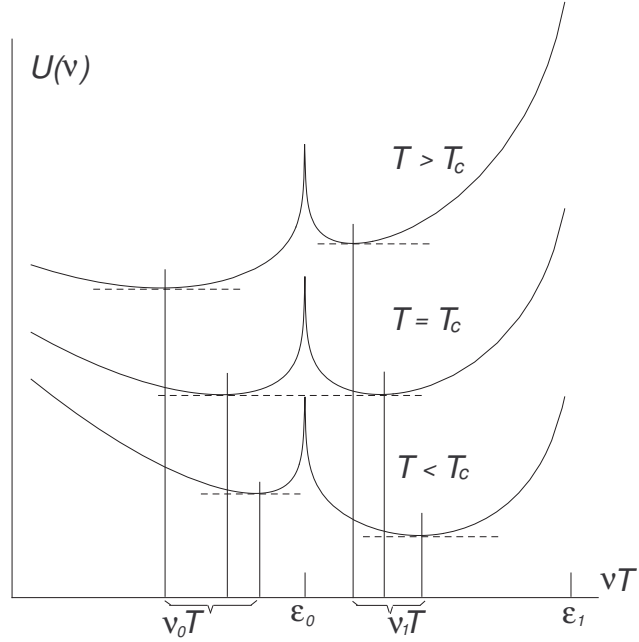} \caption{\small
Behavior of the function $U(\nu)$ in the vicinity of $\nu=\varepsilon_{0}/T$.}
\end{center} \end{figure}

We denote by $U(-ix)$ the exponent of the integrand in Eq.~(\ref{9a}):
\be\label{32a} U(-ix)=ixN+W(-ix).\ee
It follows from the definition of $W(\nu)$ [see Eq. (\ref{10a})] that the function $U(-ix)$ is singular
at the points $x_{l}=i\varepsilon_{l}/T$, and its real part goes to infinity at these points. Of course, the
function $U(-ix)$ reaches its minimum value at points $i\nu_{l}$ lying along the imaginary axis
between each pair of singular points:
$$U'(\nu_{l})=0,\quad
\frac{\varepsilon_{l-1}}{T}<\nu_{l}<\frac{\varepsilon_{l}}{T}.$$
The behavior of the function $U(\nu)$ in the vicinity of the first singularity $x_{0}$ is shown schematically
in Fig. 2.

The depth of the minimum of the function $U(\nu)$ at the saddle points
$\nu_{0}$ and $\nu_{1}$ depends on temperature: $U(\nu_{0})<U(\nu_{1})$
for $T>T_{c}$ and, oppositely, $U(\nu_{0})>U(\nu_{1})$ for $T<T_{c}$.
Therefore for an optimal estimate of the integral (\ref{9a}) for $T<T_{c}$
the integration contour must be deformed so that it passes through the
saddle point $x_{s}=i\nu_{1}$, as shown in Fig. 3. At the point $x_{0}$
the function $\exp[U(-ix)]$ has a simple pole. The contribution from this
pole to the integral (\ref{9a}), which we denote $Z_{0}$, is equal to the
residue of the integrand there:
 \be\label{33a} \ln
Z_{0}=-\frac{N\varepsilon_{0}}{T}+V\biggl(\frac{\varepsilon_{0}}{T}\biggr),\ee
where the divergence $V(\nu)$ is defined in Eq. (\ref{22a}). The
contribution of the integral along the contour $C$ has the form
 \be\label{34a}
Z_{1}=\frac{1}{2\pi}\int\limits_{C}dx\,\e^{U(-ix)}\simeq\frac{\e^{U(\nu_{1})}}{\sqrt{2\pi
U''(\nu_{1})}}\,.\ee

\begin{figure}[htb]
\begin{center}
\includegraphics[height=70mm,keepaspectratio=true]
{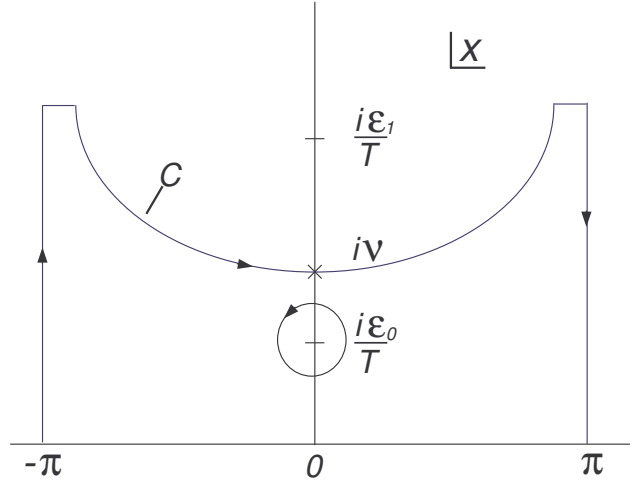} \caption{\small
Integration contour for the integral in Eq. (\ref{9a}) for $T<T_{c}$.}
\end{center} \end{figure}

We note that, as a consequence of the periodicity of the function $W(\nu)$, i.e.,
$$W(\nu+2\pi i)=W(\nu),$$
the contributions from the parts of the integration contour along the ray
$[-\pi,\,-\pi+i\infty)$ and $[\pi,\,\pi+i\infty)$, being equal in
magnitude and opposite in sign, cancel each other out. Furthermore, it is
easy to see from expressions (\ref{33a}) and (\ref{34a}) that for
$T<T_{c}$ and $N\gg1$ the ratio of the integrals $Z_{1}/Z_{0}$ is
exponentially small, and therefore the partition function in the BEC
regime is given by the exceedingly simple expression (\ref{33a}). It
follows from that expression, in particular, that the entropy
$$S=\frac{\partial\ln(T\ln Z_{0})}{\partial
T}=\sum_{k=1}^{\infty}[\ln(\widetilde{n}_{k}+1)+
\widetilde{n}_{k}(\varepsilon_{k}-\varepsilon_{0})/T]\,,$$
goes to zero at $T\to0$, as it should.

\section{Modified Asymptotic Expansion\\ of the Partition Function}
The most interesting region, but the hardest for calculations, is the
critical neighborhood of $T_{c}$. Here the contributions of $Z_{0}$ and
$Z_{1}$  are of the same order of magnitude, and the fluctuations are
maximal. To obtain the correct asymptotic expansion of the integral
(\ref{9a}) in inverse powers of the number of particles $N$ we propose the
following approach, consisting of several steps.

{\it i}) In the first step we separate out explicitly the first singular term in the integrand of
(\ref{9a}):
\be\label{35a}
Z=\frac{1}{2\pi}\int\limits_{-\pi}^{\pi}\frac{dx\
\e^{U(-ix)}}{2\sh\bigl(\frac{\varepsilon_{0}}{2T}+
\frac{ix}{2}\bigr)}\,,\ee
where, in contrast to (\ref{32a}), the function U(-ix) here has a different form:
 \be\label{36a}
U(-ix)=\frac{\varepsilon_{0}}{2T}+ix(N+\Pd)+V(-ix).\ee
The saddle point $x_{s}=i\nu$ for the function (\ref{36a}) satisfies the equation [cf. Eq. (\ref{14a})]
 \be\label{37a} V'(\nu)=N+\Pd.\ee

{\it ii}) In the second step we make the change of integration variable $x\to u$:
 \be\label{38a}
u^{2}=U(\nu)-U(-ix)=V(\nu)-V(-ix)-(\nu+ix)(N+\Pd),\ee \be\label{39a}
2u\,du=iU'(-ix)=i[V'(-ix)-N-\Pd].\ee

{\it iii}) Finally, we deform the integration contour in the $x$-plane so that it passes through
the saddle point along the line of steepest descent, which is determined by the equation
$$\Ima[U(-ix)]=0.$$
As a result of these steps we can transform the integral in (\ref{35a}) to a form in which
  \bee\label{40a}
Z&=&Z_{0}+Z_{1},\\\label{41a}
Z_{0}&=&\frac{\e^{U(\nu)}}{2\pi}\int\limits_{-\infty}^{\infty}\frac{du\
\e^{-u^{2}}}{v+iu}\,,\\\label{42a}Z_{1}&=&\frac{\e^{U(\nu)}}{2\pi}\int\limits_{-\infty}^{\infty}du\
\e^{-u^{2}}f(u),\eee
and denote by $iv$ the value of the variable $u$, defined in Eq. (\ref{38a}), corresponding to the point
$x=i\varepsilon_{0}/T$ :
 \be\label{43a}
v^{2}=U(\tfrac{\varepsilon_{0}}{T})-U(\nu)=V(\tfrac{\varepsilon_{0}}{T})-V(\nu)-
(\tfrac{\varepsilon_{0}}{T}-\nu)(N+\Pd).\ee

The function $f(u)$ in the integrand of (\ref{42a}) is analytic in the
neighborhood of the point $u=iv$: \be\label{44a}
f(u)=\frac{u}{iU'[-x(u)]}\,
\frac{1}{\sh[\frac{\varepsilon_{0}}{2T}+\frac{ix(u)}{2}]}-\frac{1}{v+iu}\,,
\ee where the function $x(u)$ in (\ref{44a}) is determined by Eq.
(\ref{38a}). The integral in (\ref{41a}) can be expressed in terms of the
error function: \be\label{45a}
Z_{0}=\Pd\,\e^{U(\nu)+v^{2}}\Erfc(v)=\Pd\,\e^{V(\varepsilon_{0}/T)-N\varepsilon_{0}/T}\Erfc(v),\ee
which satisfies the well-known equations \cite{GrRy} \be\label{46a}
\Erfc(v)=\frac{2}{\sqrt{\pi}}\int\limits_{v}^{\infty}dx\ \e^{-x^{2}},\quad
\Erfc(0)=1,\quad \Erfc(-\infty)=2,\quad \Erfc(-v)=2-\Erfc(v),\ee and, for
$v\gg1$, it has the asymptotic expansion \be\label{47a}
\Erfc(v)=\frac{\e^{-v^{2}}}{v\sqrt{\pi}}\biggl(1-\frac{1}{2v^{2}}+\frac{3}{4v^{4}}+
O(v^{-6})\biggr)\,.\ee

We find an approximate value of the integral (\ref{42a}) by expanding the function (\ref{44a}) in a
Taylor series at the point $u=0$:
 \be\label{48a}
Z_{1}=\frac{\e^{U(\nu)}}{2\sqrt{\pi}}[f(0)+\tfrac{1}{4}f''(0)+\cdots].\ee
It follows from Eq. (\ref{39a}) that $dx/du=-2iu/U'(ix)$. Ultimately we obtain for the function
$f(u)$ and its second derivative at the point $u = 0$
\bee\nonumber
f(0)&=&\frac{1}{\sqrt{2V''(\nu)}\sh(\frac{\varepsilon_{0}}{2T}-\frac{\nu}{2})}-\frac{1}{v}\,,\\
\label{49a} f''(0)&=&\frac{2}{v^{3}}-\frac{2}{[2V''(\nu)]^{3/2}
\sh(\frac{\varepsilon_{0}}{2T}-\frac{\nu}{2})}\times\\\nonumber&&\times
\biggl[
\frac{1}{\sh^{2}(\frac{\varepsilon_{0}}{2T}-\frac{\nu}{2})}-\frac{V^{'''}(\nu)}{V^{''}(\nu)}
\coth(\frac{\varepsilon_{0}}{2T}-\frac{\nu}{2})+\frac{5}{6}
\biggl(\frac{V^{'''}(\nu)}{V^{''}(\nu)}\biggr)^{2}-
\frac{V^{(4)}(\nu)}{2V''(\nu)}+\frac{1}{2}\biggr].\eee In the limit $T\to
T_{c}$ we have $\nu\to \varepsilon_{0}/T$, $v\to 0$. Then, resolving the
uncertainty $(\infty\, -\infty)$ at the point $v = 0$, we obtain from
(\ref{49a}) for the crossover region
 \be\label{50a}
\begin{array}{l}\displaystyle f(0)=\frac{V^{(''')}(\nu)}
{3\sqrt{2}[V''(\nu)]^{3/2}}\,,\\\ \\\displaystyle
f''(0)=\frac{V^{'''}(\nu)}{3\sqrt{2}[V''(\nu)]^{5/2}}\biggl[1-\frac{35}{9}
\biggl(\frac{V^{'''}(\nu)}{V''(\nu)}\biggr)^{2}+\frac{5V^{'''}(\nu)}{V''(\nu)}-
\frac{6V^{(5)}(\nu)}{5V^{'''}(\nu)}\biggr]\,.\end{array}\ee The
contributions $f(0)$, $f''(0)$, and their ratio $f''(0)/f(0)$ all fall off
with increasing $N$. The power of the decrease of the ratio $f''(0)/f(0)$
depends on the asymptotic behavior of the derivatives $V^{(p)}(\nu)$,
which, in turn, is determined by the concrete form of the spectrum
$\varepsilon_{l}$ or the spectral density $g_{l}$ as functions of $l$. For
the quadratic confining potential considered in the Appendix, the
derivatives $V^{(p)}\sim N$ for $p<3$, $V^{(3)}\sim N\ln N$ and
$V^{(p)}\sim N^{p/3}$ for $p>3$. It follows from this that representations
(\ref{45a}) and (\ref{48a}) for the contributions to the partition
function are indeed asymptotic expansions in inverse powers of $N$ for all
temperatures, including the critical temperature region $T\sim T_{c}$.

As we see from Figs. 2 and 4, even at relatively small numbers of particles (as low as
$N\sim100$) the discrepancy between the exact expressions for the fluctuations of the Bose
condensate and the approximate expressions corresponding to the representation of the
partition function in Eqs. (\ref{40a}), (\ref{45a}), (\ref{48a}), and (\ref{49a}) is indiscernable on the graphs. It can be
shown that at a temperature below $T_{c}$ the given representation coincides with the asymptotic
expression (\ref{33a}), and for $T>T_{c}$ it goes over to (\ref{13a}).

\begin{figure}[htb]
\begin{center}
\includegraphics[height=100mm,keepaspectratio=true]
{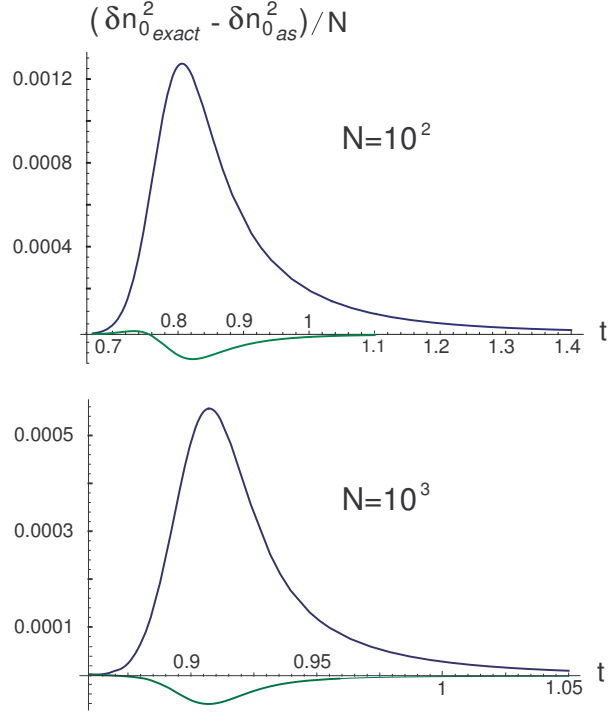} \caption{ \small The difference between the approximate
expressions for the fluctuations and the exact values $(\delta \n_{0\,
exact}^{2}-\delta \n_{0\, as}^{2})/N$. Curve 1 corresponds [see Eq.
(\ref{49a})] to the leading asymptotic contribution $f(0)$, curve 2
includes the asymptotic correction $f''(0)$.}
\end{center} \end{figure}
\begin{figure}[htb]
\begin{center}
\includegraphics[height=100mm,keepaspectratio=true]
{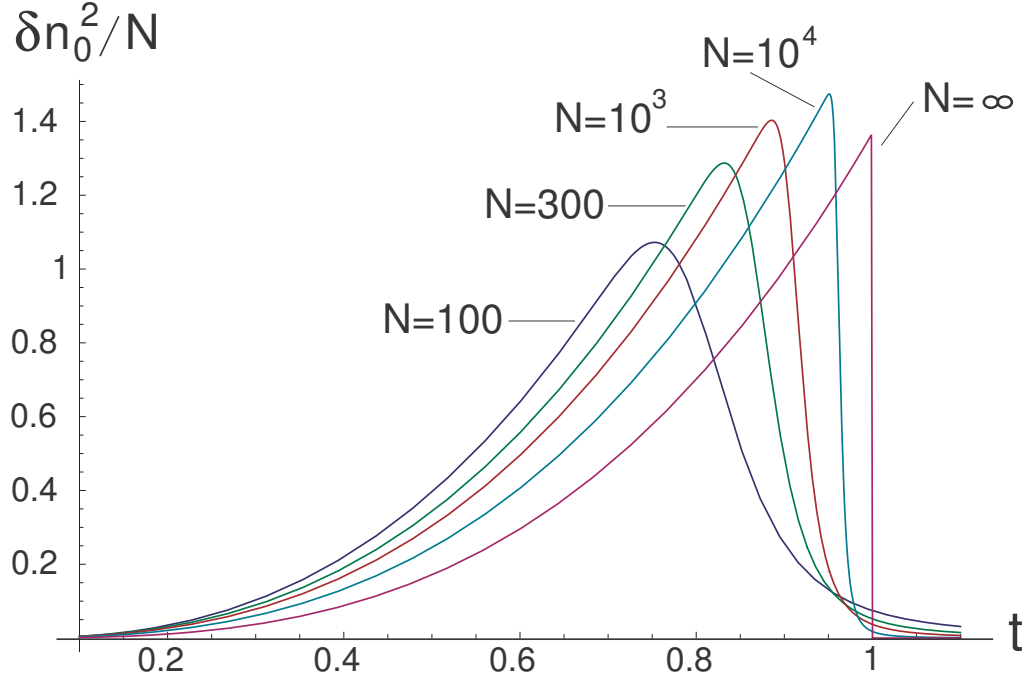} \caption{ \small Dependence of the relative fluctuation $\delta
n_{0}/N$ on temperature $t=T/T_{BEC}$ for different values of the number
of particles in the system.}
\end{center} \end{figure}
The critical region is determined by the condition \be\label{51a}
|\Erfc(v)-1|\leqslant\epsilon, \quad \epsilon\ll1.\ee For example, setting
$\epsilon=10^{-1}$, we find from (\ref{51a}) the boundary value
$v_{\epsilon}=1.16$ and, using the asymptotic approximation for $V(\nu)$
[see Eq. (\ref{21p}) in the Appendix], we obtain for the critical
temperature region $|T-T_{c}|\leqslant\Delta T$
 \be\label{52a} \frac{\Delta
T}{T_{\BEC}}=\frac{\Delta\tau}{\Delta\tau_{\BEC}}=v_{\epsilon}\biggl(\frac{2}{\zeta(2)}\biggr)^{1/2}
\biggl(\frac{\zeta(3)}{N}\biggr)^{1/3}\simeq\frac{1}{N^{1/3}} .\ee
It is seen from Eq. (\ref{52a}) that this region narrows extremely slowly with increasing particle
number. In other words, the thermodynamic limit is reached at very large particle numbers
$N\gtrsim10^{6}$. Figure 5 illustrates the evolution of the relative fluctuation of the density of the
Bose condensate as the total number of particles in the system is varied. The curves shown
convincingly demonstrate that even for a system with $N\approx10^{4}$ the temperature behavior of
these fluctuations nevertheless differs markedly from the limit $N\to\infty$, in which
\be\label{53a} \frac{\delta
\n_{0}^{2}}{N}=\frac{V''(\varepsilon_{0}/T)}{N}\,\theta(T_{\BEC}-T)=\frac{\zeta(2)}{\zeta(3)}
\biggl(\frac{T}{T_{\BEC}}\biggr)^{3}\theta(T_{\BEC}-T).\ee

\section{Conclusion}
As we have said, in the standard theoretical treatment of the BEC problem the trap is a
three-dimensional box with, importantly, a fixed volume. In this case the Bose condensate
formation temperature (on the quantum scale $\tau=T/\hbar\omega_{\BOX}$) has the form
\be\label{1b}
\tau_{\BEC}^{\BOX}=\biggl(\frac{N}{\zeta(3/2)}\biggr)^{2/3}\,,\ee
while in the problem considered above, for a trap in which the confining potential is quadratic
and the volume is not fixed,
 \be\label{2b}
\tau_{\BEC}^{\Trap}=\biggl(\frac{N}{\zeta(3)}\biggr)^{1/3}\,.\ee
As is seen from Eqs. (\ref{1b}) and (\ref{2b}), the dependence of the BEC temperature on the number
of particles in the system is significantly different in these two cases. For the box the
characteristic energy is expressed in terms of the volume V of the system and the mass $M$
of the particle in the following way:
 \be\label{3b}
\hbar\omega_{\BOX}=\frac{\pi^{2}\hbar^{2}}{2\,{\text{V}}^{2/3}M}\,.\ee
In the magnetic trap the volume is not fixed and, moreover, it changes with changing temperature.
We define the effective size of the system in a trap with a quadratic confining potential
as the amplitude of the oscillations of an oscillator with energy equal to the temperature $T$.
Then
\be\label{4b}
{\text{V}}^{eff}\simeq\frac{4\pi}{3}\left(\frac{T}{M\omega^{2}}\right)^{3/2}\,.\ee
Now, after expressing the number of particles in terms of the particle density $\rho$ ($N=\rho$V)
and substituting Eq. (\ref{3b}) into (\ref{1b}) and  (\ref{4b}) into (\ref{2b}), we find for $T_{\BEC}$ in the two cases
\be\label{5b}\begin{array}{l}\displaystyle
T_{\BEC}^{\BOX}=\frac{\pi^{2}\hbar^{2}}{2M}\biggl(\frac{\rho}{\zeta(3/2)}\biggr)^{2/3}\,,\\\
\\\displaystyle
T_{\BEC}^{\Trap}=\frac{\hbar^{2}}{M}\biggl(\frac{4\pi\rho}{3\zeta(3)}\biggr)^{2/3}\,.\end{array}\ee
It follows from (\ref{5b}) that in the language of particle number density the the temperatures of
BEC in the box and trap are close not only qualitatively but also quantitatively:
\be\label{6b} \frac{T_{\BEC}^{\BOX}}{T_{\BEC}^{\Trap}}=\frac{1}{2}
\biggl[\frac{3\pi^{2}\zeta(3)}{4\zeta(3/2)}\biggr]^{2/3}\simeq1.13.\ee

On the whole, it should be emphasized one again that BEC is rightfully considered to be
one of the fundamental discoveries of theoretical physics. Its clearest trait is not simply the
accumulation of a macroscopic number of particles in the ground state upon cooling of an
ideal Bose gas but the fact that this process is a phase transformation in a system of mutually
{\it noninteracting} particles. Most likely the term ``Bose condensation'' came into use because
of the analogy (which, strictly speaking, is not entirely correct) with the condensation of a
vapor to a liquid, which was discussed by Einstein in his pioneering paper \cite{1}.

Although phase transformations in nature are extremely diverse, at the same time they
demonstrate surprising universality: the change of the thermodynamic properties of a system
occurs abruptly when the temperature (or some other controllable parameter) crosses its
critical value. From a formal theoretical physics point of view the main question is, how,
in functions which are initially analytic in temperature, does the singularity arise at the
critical point $T_{c}$:
 \be\label{6bb}
f(T,N)\mathop{=}_{N\to\infty}f_{1}(T)\,\theta(T_{c}-T)+f_{2}(T)\,\theta(T_{c}-T).\ee
Bose-Einstein condensation, as an exactly solvable model, gives a simple answer to this
question. For example, in the GCE the role of the ``smeared'' $\theta$-function, according to Eq.
(\ref{12}), is played by the quantity
\be\label{7bb} \theta_{\text{GCю}}(x)=\frac{1}{2}
\left(\frac{x}{\sqrt{x^{2}+\alpha_{\text{GCю}}/N}}+1\right),\qquad
\alpha_{\text{GCю}}=\frac{4\zeta(2)}{\zeta(3)}\,,\ee
and in the CE, as follows from Eq. (\ref{45a}), by
 \be\label{8bb}
\theta_{\text{Cю}}(x)=\frac{1}{2}\Erfc(-\alpha_{\text{Cю} }\sqrt{N}\,
x),\qquad \alpha_{\text{Cю}}=\sqrt{\frac{\zeta(3)}{2\zeta(2)}}.\ee
For both representations (\ref{7bb}) and (\ref{8bb}) at $N\to\infty$ the limit is the ordinary $\theta$-function, but
at a finite value of $N$ they behave differently. It is seen in Fig.~6 that $\theta_{\text{Cю}}$ is closer to a step
than $\theta_{\text{GCю}}$. Thus one can say that the thermodynamic limit sets in somewhat faster in the
CE than in the GCE.

\begin{figure}
[htb]
\begin{center}
\includegraphics[height=90mm,keepaspectratio=true]
{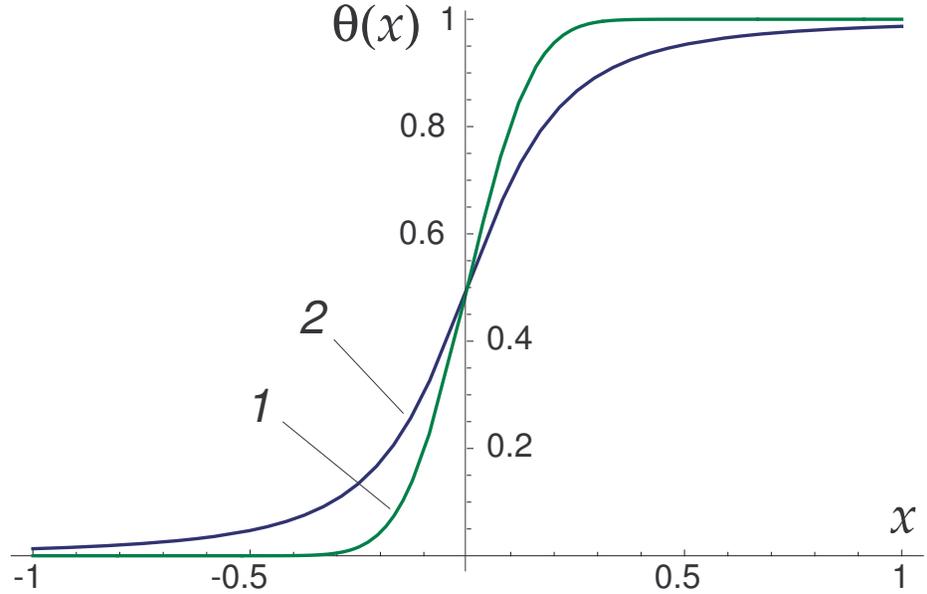} \caption{ \small
 Behavior of the smoothed $\theta$-functions for $N=10^{2}$. Curve 1 --- $\theta_{\text{Cю}}(x)$, curve 2 ---
$\theta_{\text{GCю}}(x)$ .}
\end{center} \end{figure}

If as the function $f(T,N)$ in (\ref{6bb}) we take the specific free energy
$$f(T,N)=-\frac{T}{N}\ln Z,$$
then both $f_{1}(T)$ and $f_{2}(T)$ are identical in the two ensembles, thus implying the equivalence
of the ensembles:
\be\label{9bb}f_{1}(T)=-\frac{\zeta(4)}{\zeta(3)}\,\frac{T^{4}}{T^{3}_{\BEC}}\,,\qquad
f_{2}(T)
=-\frac{\Li_{4}(\e^{\mu/T})}{\zeta(3)}\,\frac{T^{4}}{T^{3}_{\BEC}}\,.\ee
The chemical potential $\mu$ in Eq. (\ref{9bb}) as a function of temperature and particle number density
is the solution of the equation
$$T^{3}\Li_{3}(\e^{\mu/T})=\zeta(3)\,T^{3}_{\BEC}\,,$$
where $\Li_{k}(z)$ is the polylogarithm [see Appendix, Eq. (\ref{12p})].
However, the specific fluctu\-a\-ti\-ons of the Bose condensate is
different in the different ensembles: in the GCE
$$\lim_{N\to\infty}\frac{\delta \n_{0}^{2}}{N}=\left\{\begin{array}{cl}\infty&{\mbox{ОПХ}\ }T<T_{c},\\
0&{\mbox{ОПХ}\ }T>T_{c},\end{array}\right.$$
and in the CE
$$\lim_{N\to\infty}\frac{\delta \n_{0}^{2}}{N}=\frac{\zeta(2)}{\zeta(3)}\,\frac{T^{3}}{T^{3}_{\BEC}}\,\theta(T_{c}-T).$$

In and of itself the accumulation of particles at the lowest energy level
is the direct and rather trivial consequence of Bose statistics, evident
from the form of the formulas for the average occupation numbers,
Eq.~(\ref{5}). When one is talking about a phase transition, however, the
question of thermodynamic limit must also be addressed: for example, is a
number of particles $N=10^{3}$ sufficient to approach it? To speak more
precisely, the term thermodynamic limit is commonly understood to mean
letting the volume of the system go to infinity at fixed temperature with
the various densities (e.g., the particle number density $\rho$) held
constant. However, the volume, density, and temperature are dimensional
quantities: 1 meter is almost "infinite" if one is measuring in angstroms.
>From the point of view of theoretical physics, any limit should be
formulated in the language of dimensionless quantities, and in the problem
of BEC of an ideal gas there are only two such quantities: the total
number of particles in the system, $N$, and the temperature on the quantum
scale, $\tau=T/\hbar\omega$. Therefore the thermodynamic limit is, first
and foremost, $N\to\infty$, and the densities should be taken as the
ratios of the corresponding extensive quantities to the total number of
particles. The asymptotic corrections to the limit $N\to\infty$ in the our
problem are of the order of $N^{-1/3}$, so that, it would seem, the
thermodynamic limit is reached with $10\%$ accuracy if $N=10^{3}$. On the
other hand, however, the BEC temperature for this number of particles is
comparable to the ground state energy $\varepsilon_{0}$ (we recall that in
the given problem $T_{c}\simeq6\varepsilon_{0}$ at $N=10^{3}$), and so one
cannot speak of a macroscopic scale of the physical quantities.

In this regard we note that when atoms of alkali metals are held in magnetic traps the
procedure of preparing a coherent state of $N$ particles is said to involve ``cooling'' apparently
to reflect its thermodynamic nature. Meanwhile, neither the volume nor the temperature
nor, moreover, the spectral density of the particle number cannot be controlled to the required
precision because of technical shortcomings and for fundamental reasons: what is
the temperature on quantum scales $T\sim \varepsilon_{0}$? Therefore, in light of the results presented
above, there is ample justification for concluding that the assertion that true BEC has been
observed in these undeniably outstanding experiments should be taken with a degree of
caution.

This study was supported in part by grants from SCOPES SNSF, the
French-Ukrainian Program ``Dnepro''-project M/185-2009, PICS CNRS and the
National Academy of Sciences of Ukraine 2009-2011, NAS Ukraine programs
"Nanostructural systems, nanomaterials, nanotechnology" - 10/7-N, and also
a target program of the Division of Physics and Astro\-no\-my NAS Ukraine.

\setcounter{equation}{0}
\renewcommand{\theequation}%
{{\mbox{A}}.\arabic{equation}}
\section*{Appendix}

The function $W(\nu)$  (\ref{10a}) and its derivatives $W^{(p)}(\nu)$ is determined by series of the form
$\sum\limits_{l=0}^{\infty}h(l)$. Separating $m$ the first m terms, we denote the remainder of the series as
 \be\label{1p}
K=\sum_{l=m+1}^{\infty}h(l)\ee
and take an Abel-Plana transform of it. Then the series in Eq. (\ref{1p}) is transformed to the
sum of two integrals,
\be\label{2p} K=I+J,\ee
in which
\bee\label{3p}
I&=&\int\limits_{m+\Pd}^{\infty}dl\, h(l)\,,\\
\label{4p}J&=&-2\int\limits_{0}^{\infty}\frac{dx\,\Ima[h(m+\Pd-ix)]}
{\e^{2\pi x}+1}\,.\eee
In order for series (\ref{1p}) to converge, the function $h(l)$ must fall off with increasing $l$ not
slower than $l^{-1}$. For a power-law function $h(l)\sim l^{-\alpha}$ its derivatives $h^{(p)}(l)$ behaves at large $l$
as $l^{-\alpha-p}$. In this case the integral $J$ can be evaluated with the aid of an asymptotic expansion
in inverse powers of $m$. We expand the function $h(m+\Pd-ix)$ in the integrand of (\ref{4p})
in a Taylor series at the point $x = 0$. For the sum of any finite number $n$ of terms in this
expansion one can switch order of the summation and integration and get
 \be\label{5p} J\simeq
\sum_{l=0}^{n}(-1)^{l}c_{l}h^{(2l+1)}(m+\Pd),\ee where \be\label{6p}
c_{l}=\frac{1}{(2l+1)!}\int\limits_{0}^{\infty}\frac{dx\,
x^{2l+1}}{\e^{2x}+1}=\frac{(1-2^{-2l-1})\zeta
[2l+2]}{(2\pi)^{2(l+1)}}\,.\ee
We note that the optimal number of terms in the asymptotic expansion (\ref{5p}) is determined
by the form of the coefficients $c_{l}$. Therefore evaluation of the integral $J$ can be done with
the aid of (\ref{5p}) to any desired accuracy by increasing $m$.

We now use this technique to evaluate the logarithm of the partition function (\ref{2a}). Here
the function $h(l)$ has the form
 \be\label{7p}
h(l)=g(l)\ln(n_{l}+1),\ee \be\label{8p} g(l)=\frac{1}{2}(l+1)(l+2),\quad
n_{l}=\frac{1}{\e^{(l-\sigma)/\tau}-1}\,,\quad
\tau=\frac{T}{\hbar\omega}\,,\quad \sigma=\nu\tau-\frac{3}{2}\,. \ee Then
\be\label{9p} W(\nu)=\sum_{l=0}^{m}g(l)\ln(n_{l}+1)+I+J.\ee
The integral $I$ in (\ref{9p}) is expressed in terms of the polylogarithms (Lerch functions)
 \be\label{10p}
I=\tau^{3}\Li_{4}(z)+a\tau^{2}\Li_{3}(z)+b\tau\Li_{2}(z),\ee where
$$a=g'(m+\Pd)=m+2,\quad
b=g(m+\Pd)=\Pd(m+\tfrac{3}{2})(m+\tfrac{5}{2}),$$ \be\label{11p}
z=\e^{-(m+\Pd-\sigma)/\tau},\ee \be\label{12p}
\Li_{k}(z)=\frac{1}{\Gamma(k)}\int\limits_{0}^{\infty}\frac{dx\,x^{k-1}}{\e^{x}/z+1}=
\sum_{l=1}^{\infty}\frac{z^{l}}{l^{k}}\,,\quad
\Li_{k-1}(z)=z\frac{d\,\Li_{k}(z)}{dz}\,.\ee
The contribution $J$ in (\ref{9p}) has the form
\be\label{13p} J=\sum_{l=0}^{n}(-1)^{l}c_{l}f_{l}\,,\ee where
\be\label{14p}f_{0}=a\ln(r+1)-b\tau^{-1}r,\ee for $l>0$ \be\label{15p}
f_{l}=-l(2l+1)\tau^{-2l+1}r_{2l-1}+a(2l+1)\tau^{-2l}r_{2l}-b\tau^{-2l-1}r_{2l+1},\ee
$$r_{0}=r=n_{m+\frac{1}{2}}=\frac{1}{\e^{(m+\frac{1}{2}-\sigma)/\tau}-1}\,,\quad
r_{l}=\tau^{l}\frac{\partial^{l}r}{\partial\sigma^{l}}=r(r+1)\frac{\partial
r_{l-1}}{\partial r}\,.$$

For calculating the asymptotic expansion of $W(\nu)$ at large $\tau$ it is sufficient to keep only
one term in the sum on the right-hand side of (\ref{9p}): ($m = 0$ in (\ref{1p})). Then, using the
well-known asymptotic expansion of the polylogarithm for  $x\to0$ \be\label{16p}
\Li_{4}(\e^{-x})=\zeta(4)-\zeta(3)x+\zeta(2)\frac{x^{2}}{2}+\biggl(\ln
x-\frac{11}{6}\biggr)\frac{x^{3}}{6}+O(x^{4}),\ee where  $\zeta(j)$ is the Riemann $\zeta$-function
 $$\zeta(4)=\frac{\pi^{4}}{90},\quad
\zeta(3)\simeq1.202,\quad \zeta(2)=\frac{\pi^{2}}{6}\,,$$
with accuracy up to terms that do not fall off with increasing $\tau$ we find that
\be\label{17p}\begin{array}{l}
I=\frac{\pi^{4}}{90}\tau^{3}+g'(\sigma)\zeta(3)\tau^{2}+g(\sigma)\frac{\pi^{2}}{6}\tau+\\\
\\\ \ \ +\frac{1/2-\sigma}{6}\left\{\bigl[g(\sigma+1)+\frac{5}{2}\bigr]
\ln\bigl[\frac{1/2-\sigma}{\tau}\bigr]-\frac{1}{6}\bigl[11(\sigma+2)^{2}-\sigma-\frac{3}{4}
\bigr]\right\}+O(\tau^{-1}).
\end{array}\ee
Here the contribution (\ref{4p}) takes the form
\be\label{18p}\begin{array}{l}\displaystyle
J=2\int\limits_{0}^{\infty}\frac{dx}{\e^{2\pi
x}+1}\Ima\bigl[g(\tfrac{1}{2}-ix)\ln(1-\e^{-(\Pd-\sigma-ix)/\tau})\bigr]\simeq\\\
\\\ \ \displaystyle\simeq\frac{\ln\tau}{12}+2\int\limits_{0}^{\infty}\frac{dx}{\e^{2\pi
x}+1}\Ima\bigl[g(\tfrac{1}{2}-ix)\ln(\tfrac{1}{2}-\sigma-ix)\bigr]+O(\tau^{-1}).\end{array}\ee
The asymptotic expansion for the derivatives $W^{(p)}(\nu)$ we obtain by
differentiating expres\-si\-ons (\ref{10p}) and (\ref{18p}) with respect
$\sigma$:
$$W^{(p)}(\nu)=\frac{\tau^{p}\partial^{p}}{\partial\sigma^{p}}\bigl[\ln(n_{0}+1)+I+J\bigr].$$
We note that, starting with the third derivative, expression (\ref{10p}) for $I$ can be written in
terms of elementary functions, since
\be\label{19p}
\Li_{1}(\e^{-x})=-\ln(1-\e^{-x}).\ee
The derivative $\partial
J/\partial\sigma$ is expressed in terms of the special function  $\psi(z)=\Gamma'(z)/\Gamma(z)$:
 \be\label{20p} \frac{\partial
J}{\partial\sigma}=-2\int\limits_{0}^{\infty}\frac{dv}{\e^{2\pi
x}+1}\Ima\biggl[\frac{g(\frac{1}{2}-ix)}{\frac{1}{2}-\sigma-ix}\biggr]=\frac{1}{48}-g(\sigma)
\bigl[\psi(1-\sigma)-\ln(\Pd-\sigma)\bigr]\,.\ee
In particular, for the derivative $W'(\nu)=\sum\limits_{l=0}^{\infty}g(l)n_{l}$
 with Eq. (\ref{20p}) taken into account, we
arrive at the expression
 \bee\nonumber W'(\nu)&=&
n_{0}+\zeta(3)\tau^{3}+g'(\sigma)\frac{\pi^{2}}{6}\tau^{2}+\tau
g(\sigma)\bigl[\ln\tau-\psi(1-\sigma)\bigr]+\\\label{21p}
&&+\frac{\tau}{4}\bigl(3\sigma^{2}+5\sigma-\tfrac{19}{6}\bigr)+\frac{1}{6}\bigl(\Pd-\sigma\bigr)
\bigl[g(\sigma+1)+\tfrac{5}{4}\bigr]-\frac{1}{24}+O(\tau^{-1}).\eee

\renewcommand{\refname}{References}

\end{document}